\documentclass[prl,twocolumn,showpacs,superscriptaddress]{revtex4}
\usepackage{dcolumn}
\usepackage{bm}
\usepackage{graphicx}
\usepackage{amsmath}
\usepackage{amssymb}
\usepackage{bm}
\begin{document}
\title{Analysis of the normal state magnetotransport in CeIrIn${_5}$}
\author{Sunil Nair}
\affiliation{Max Planck Institute for Chemical Physics of Solids,
Noethnitzer Str. 40, 01187 Dresden, Germany}
\author{M.~Nicklas}
\affiliation{Max Planck Institute for Chemical Physics of Solids,
Noethnitzer Str. 40, 01187 Dresden, Germany}
\author{J.~L.~Sarrao}
\affiliation{Los Alamos National Laboratory, Los Alamos, New
Mexico 87545, USA}
\author{J.~D.~Thompson}
\affiliation{Los Alamos National Laboratory, Los Alamos, New
Mexico 87545, USA}
\author{F.~Steglich}
\affiliation{Max Planck Institute for Chemical Physics of Solids,
Noethnitzer Str. 40, 01187 Dresden, Germany}
\author{S.~Wirth}
\affiliation{Max Planck Institute for Chemical Physics of Solids,
Noethnitzer Str. 40, 01187 Dresden, Germany}
\date{\today}
\begin{abstract}
We present an analysis of the normal state magnetotransport in the
heavy fermion superconductor CeIrIn${_5}$. The Hall effect and the
transverse magnetoresistance in this material do not appear to be
uniquely correlated, as inferred from the field dependence of the
current ratio ($R_{\sigma} = \sigma_{xy} / \sigma_{xx}^2 H$). The
Hall coefficient is seen to satisfy a scaling equation of the form
$R_H = f [H / (a + b \, T^c)]$. These results are compared to
those observed earlier in CeCoIn$_5$, and are discussed in terms
of the contrasting phase diagram which the CeIrIn$_5$ system
exhibits in relation to its Co counterpart.
\end{abstract}
\pacs{Heavy Fermion superconductors, Hall effect,
Magnetoresistance} \maketitle

\hspace*{-0.3cm}{\bf I. INTRODUCTION}\\[0.3cm]
The phenomenon of superconductivity in the heavy fermion metals
remains an area of extensive theoretical and experimental
research. The initial interest in this phenomenon was focused on
understanding how a superconducting condensate could form in a
regime where magnetic fluctuations are known to be dominant, since
superconductivity and magnetism were thought to be antithetical to
each other. This apparent contradiction is now thought to be
lifted by the fact that in these materials, the bosonic mode which
facilitates Cooper pair formation may be the incipient
antiferromagnetic fluctuations itself \cite{mathur}. This
fascinating interplay between magnetism and superconductivity has
received further impetus due to the often observed existence of
superconductivity in the vicinity of Quantum Critical Points
(QCP)---continuous quantum phase transitions (QPT) at absolute
zero temperature driven by an external control parameter. In the
heavy fermion metals, these extraordinary transitions arise as a
consequence of the competition between two fundamental processes:
the Ruderman-Kittel-Kasuya-Yosida (RKKY) interaction which
promotes a magnetically ordered ground state, and the Kondo
effect, which shields the local moments.

In recent years, the Ce$M$In$_5$ ($M =$ Co, Ir or Rh) family of
compounds has emerged as a fertile playground where many of these
competing interactions can be individually tailored \cite{sar}.
Not surprisingly, this manifests itself in the form of an
extremely rich and varied phase diagram. For instance, though
CeCoIn$_5$ and CeIrIn$_5$ are both ambient pressure
superconductors, they differ drastically with respect to their
quantum critical behavior in relation to the superconductivity in
the magnetic field-temperature phase space. In CeCoIn$_5$, a
magnetic field-induced QCP is located in the vicinity of the
superconducting upper critical field ($H_{c2}$) \cite{pag},
whereas in CeIrIn$_5$ on the other hand, a QPT is speculated to be
related to a metamagnetic transition that can only be approached
by fields of the order of 25 T \cite{cap}. This marked difference
is in spite of the rather similar band structures of these
systems, as has been concluded from de Haas-van Alphen
measurements \cite{hal,hag}. These measurements (supplemented by
band structure calculations) have shown that the CeMIn$_5$ systems
have nearly cylindrical Fermi surfaces, which in turn arise as a
consequence of their quasi two dimensional crystalline structure
consisting of units of CeIn$_3$ separated by MIn$_2$ planes. This
is also manifested by a pronounced anisotropy in various physical
properties. The superconductivity in some of these systems is also
reported to be anomalous, and it has been suggested \cite{tan}
that in CeCoIn$_5$ a group of conduction electrons may not
participate in the formation of the superconducting condensate.
This can be looked upon as a rather unique form of electronic
phase separation, and---if reconfirmed by other
measurements---would represent the most extreme case of multiband
superconductivity.

Current interest in these systems is focused on not only
unraveling the rich phase diagrams which these systems exhibit,
but is also stemming from the fact that many of the physical
properties---both in the normal and superconducting state---are
remarkably similar to those exhibited by the high-temperature
superconducting cuprates \cite{nak}. For instance in the CeMIn$_5$
systems, the resistivity is known to have a linear temperature
dependence, and the Hall coefficient is strongly temperature
dependent. The superconducting gap function has line nodes and has
been suggested to be of $d$-wave symmetry \cite{mat}. Moreover,
the formation of the superconducting condensate appears to be
preceded by a precursor state, in similarity to the pseudogap
state in the cuprates. Here, we report on the analysis of the Hall
effect and magnetoresistance in a single crystal of CeIrIn$_5$.
The correlation between the Hall and the transverse conductivity
is discussed. A scaling analysis of the Hall coefficient using a
functional form $R_H = f [H / (a + b \, T^c)]$ and its
implications
are also reported.\\[0.2cm]

\hspace*{-0.3cm}{\bf II. EXPERIMENTAL TECHNIQUES}\\[0.3cm]
All measurements are made on a single crystal of CeIrIn$_5$ with
approximate dimensions of 0.8 mm $\times$ 0.7 mm $\times$ 0.08 mm,
using the standard six-contact geometry. Simultaneous Hall effect
and magnetoresistance measurements are performed using a modified
Kelvinox-25 dilution refrigerator in the range 0.05 K $\le T \le$
2.5 K. The measurement protocol is in the form of isothermal field
sweeps, with the magnetic field $H$ up to 15 T applied parallel to
the tetragonal $c$ axis. The Hall voltages are extracted as the
asymmetric component under magnetic field reversal. Low
temperature transformers are used in conjunction with low noise
voltage preamplifiers to enable a voltage resolution of better
than $\pm$0.01 nV.\\[0.2cm]

\begin{figure}
\includegraphics[width=8.4cm,clip]{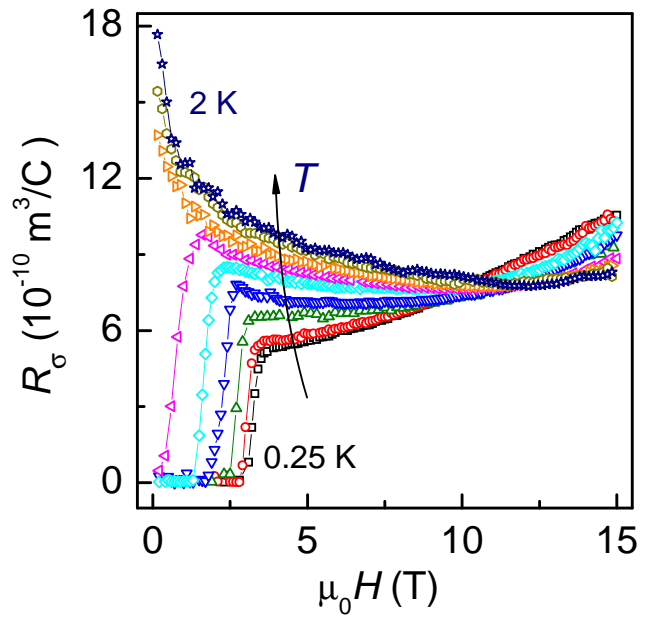}
\caption{Field dependence of the current ratio ($R_{\sigma} =
\sigma_{xy} / \sigma_{xx}^2 H$) of CeIrIn$_5$ at selected
temperatures. Unlike that reported earlier for the related system
CeCoIn$_5$, $R_{\sigma}$ is strongly field dependent implying that
the Hall and the transverse conductivities are not uniquely
correlated.} \label{fig1}
\end{figure}

\hspace*{-0.3cm}{\bf III. RESULTS AND DISCUSSION}\\[0.3cm]
A comprehension of the normal state magnetotransport of these
complex systems clearly warrants the analysis of the electrical
and Hall conductivities ($\sigma_{xx}$ and $\sigma_{xy}$,
respectively) in unison. An interesting manifestation of the
correlation between these two quantities was recently demonstrated
in CeCoIn$_5$: it was shown that the so-called current ratio
(defined as $R_{\sigma}(T,H) = \sigma_{xy} / \sigma_{xx}^2 H$) is
constant with respect to the applied magnetic field below 1 K
\cite{ono}. With this definition, the current ratio $R_{\sigma} =
R_H [1 + (\tan \theta_H)^2]$ differs from the conventionally used
$R_H (= \sigma_{xy}/H)$ at large values of the Hall angle
$\theta_H$. Here, it is to be noted that the magnitude of
$\theta_H$ which effectively measures the deflection of the charge
carriers in the material due to the applied magnetic field, is
substantially large in the Ce$M$In$_5$ systems. In CeIrIn$_5$ for
instance, $\theta_H$ attains values of the order of about
40$^{\circ}$ at applied fields of the order of 15 T \cite{nai}.
Interestingly, the constancy in $R_{\sigma}(T,H)$ was observed in
CeCoIn$_5$ in spite of the appreciable field dependence exhibited
by the individual $\sigma_{xx}(H)$ and $\sigma_{xy}(H)$
components. It was suggested that this constancy of
$R_{\sigma}(T,H)$ arises due to the fact that below 1 K the
electron mean free path ($\ell$) is significantly enhanced. Though
$R_{\sigma}$ deviates from constancy above 1 K, the application of
a magnetic field helps in recovering it to an appreciable extent.
This was proposed to occur because of the fact that the applied
magnetic field suppresses the incipient antiferromagnetic spin
fluctuations. Since spin excitations are the dominant scattering
mechanism of charge carriers, this suppression of the spin
scattering results in an enhancement of $\ell$, and consequently a
{\em correlated} increase of the diagonal ($\sigma_{xx}$) and
off-diagonal conductivities ($\sigma_{xy}$). Fig.~\ref{fig1} shows
the current ratio $R_{\sigma}$ plotted as a function of the
applied field $H$. The lack of constancy, in comparison to that
observed earlier in CeCoIn$_5$ is obvious, and $R_{\sigma}$ is
seen to be strongly field dependent.

The disparate behavior of $R_{\sigma}$ in the Co and Ir systems
may be a reflection of the difference between the low temperature
phase diagrams of these two systems: the $H$--$T$ phase space
sampled by our measurements on CeIrIn$_5$ does not encompass the
putative QPT the signatures of which have been observed in prior
investigations on CeCoIn$_5$. This disparity in $R_{\sigma}$ is,
however, in line with prior Hall effect measurements \cite{sin} in
CeCoIn$_5$ where a pressure dependent feature in the differential
Hall coefficient was observed. This feature, attributed to the
influence of critical antiferromagnetic fluctuations in
CeCoIn$_5$, was not found in CeIrIn$_5$ \cite{na2}. More
importantly, our data suggest that the field dependent excitations
responsible for quasiparticle scattering in CeIrIn$_5$ do not
appear to influence the Hall and the magnetoconductivities in a
correlated fashion. This lack of correlation between the diagonal
and off-diagonal magnetotransport quantities could arise due to
(i) the presence of hitherto unidentified excitations which act
differently on these two quantities, (ii) a non-trivial
modification in the topology of the Fermi surface, or (iii) an
anisotropy in the scattering rates along different areas of the
Fermi surface. It is to be noted that our analysis of the Hall
mobility had shown that the superconductivity in CeIrIn$_5$ is
preceded by a precursor state, similar to the pseudogap state in
the high-$T_c$ superconducting cuprates \cite{na2}. Using a single
parameter scaling of the Hall angle, it also has recently been
demonstrated that this precursor state appears to selectively
influence the Hall channel, and has relatively less influence on
the resistivity \cite{nai}.

In the temperature regime under investigation here, the Hall
coefficient $R_H$ is in itself a quantity of fundamental interest.
This is primarily fueled by the fact that at these low
temperatures the measured Hall response in heavy fermion systems
is expected to be relatively free from the influence of skew
scattering and predominantly arises only from the normal part of
the Hall effect \cite{fer}. Thus, $R_H(T,H)$ is a measure---albeit
an indirect one---of the Fermi surface volume. It has been
successfully used to investigate the evolution of the Fermi
\begin{figure}
\includegraphics[width=7.8cm,clip]{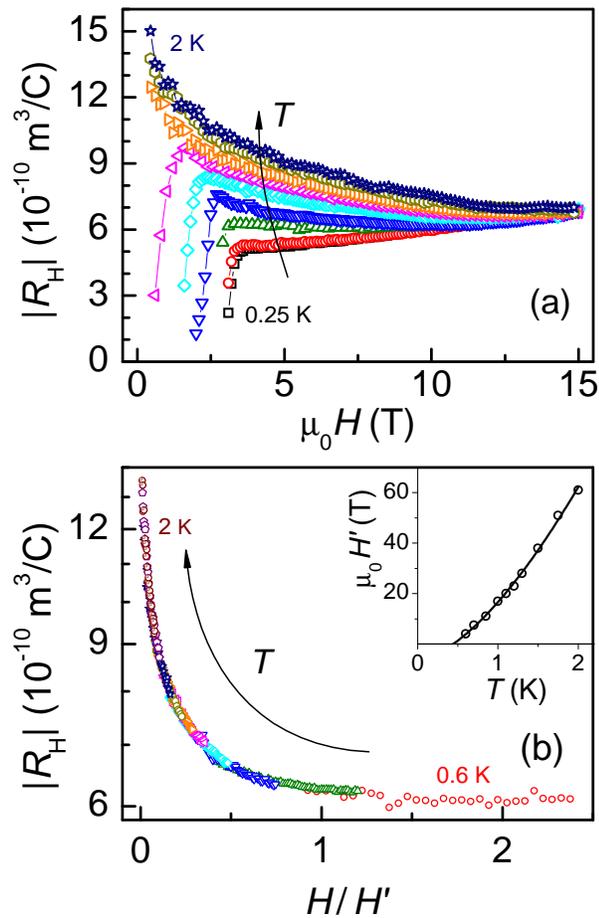}
\caption{The Hall coefficient $|R_H| = |\rho_{xy}| / H$ as
measured in CeIrIn$_5$ at selected temperatures (a) and the
scaling of the normal state $|R_H|$ achieved by normalizing the
magnetic field by $H'$ (b). The inset depicts the temperature
evolution of $H'$, and the solid line is a power law fit. }
\label{fig2}
\end{figure}
surface across a field induced quantum critical point \cite{pas}.
In the case of CeCoIn$_5$, it was shown that the measured $|R_H|$
could be scaled into a generic curve when its field dependence was
normalized by a single scaling factor \cite{sin}. This factor $H'$
(with dimension of a magnetic field) was thought to represent the
effective carrier mobility $\mu_{\rm eff}$, averaged over
different sheets of the Fermi surface which contribute to the
measured Hall voltage. $H'$ was experimentally determined by the
aforementioned feature in $R_H$ in a limited temperature range,
beyond which it was estimated by the scaling procedure.

Fig.~\ref{fig2}a depicts the $|R_H|$ vs. $H$ as measured
isothermally in CeIrIn$_5$. The sharp drop in $|R_H(H)|$
corresponds to the onset of the superconducting transition. The
normal-state scaling of $|R_H(H)|$ using a single scaling
parameter $H'$ is shown in Fig.~\ref{fig2}b where a remarkably
good overlap of experimental data is observed. It is to be noted
that, unlike in the case of CeCoIn$_5$, there is no discernable
feature corresponding to $H'$ in the raw $|R_H(H)|$ data of
CeIrIn$_5$, and here $H'$ is estimated in the whole temperature
range using the scaling procedure alone. In line with earlier
observations on CeCoIn$_5$, the temperature dependence of $H'$ can
also be reasonably fit using a power law of the form $H' = a + b\,
T^c$ (as shown in the inset) yielding $a = -6.63 \pm 2.3$ T. In
the absence of an experimental signature associated with $H'$ in
the CeIrIn$_5$ data, the extent of uncertainty in the
determination of the coefficients $b$ and $c$ (in the form of
unrecognized multiplicative factors) cannot easily be ascertained.
Hence, we limit our discussion to the value, and the implications
of, the coefficient $a$ alone. In CeCoIn$_5$ the value of $a$ as
determined from the scaling analysis was of the order of 4 T and
was suggested to be related to the existence of a QPT at that
value of the applied magnetic field. The fact that the sign of $a$
is {\em negative} in CeIrIn$_5$ possibly indicates the absence of
an antiferromagnetic QPT in the vicinity of superconductivity in
this system. Since the magnetic field $H$ influences the effective
exchange coupling between localized spins (through the RKKY
interaction), the contrast in the sign of $a$ is probably related
to an even weaker antiferromagnetic interaction in CeIrIn$_5$
compared to the Co system (see discussion below). Interestingly,
this scaling is observed to be valid only for data above 0.4 K in
the case of CeIrIn$_5$. We note that our measurements track only
the resistive transition into superconductivity which occurs at
1.2 K; the bulk transition in this system is known to be at about
0.4 K. This discrepancy remains to be fully comprehended though
some studies have suggested that this arises due to filamentary
superconductivity which is intrinsic, and involves electrons from
the part of the Fermi surface responsible for bulk
superconductivity \cite{bia}. The fact that this bulk transition
temperature is reflected in our scaling analysis---albeit in an
indirect manner---is significant. Moreover, the scaling of
$|R_H(H)|$ in CeIrIn$_5$ is also seen to be valid only within the
{\em coherent} Kondo regime of the phase diagram. This regime is
associated with a positive magnetoresistance (MR), and was
characterized by the change in the sign of
$\partial$(MR)/$\partial(H)|_T$ \cite{na2}. In the incoherent
regime the magnetoresistance is negative, since it primarily
results from the suppression of spin flip scattering. On
decreasing temperature, the scattering becomes more coherent, and
the magnetoresistance in this regime is driven by the Lorentz
force acting on the charge carriers. The fact that this single
parameter scaling is only observed in the coherent Kondo regime
clearly implies that this functional form of $R_H(T,H)$ is valid
only when the contribution that arises from scattering due to spin
fluctuations is absent.

A similar scaling form $R_H = f [H / (a + b \, T^c)]$ has been
reported to be valid for several members of the Ce$M$In$_5$
family, albeit in a different temperature regime ($T > T_c$)
\cite{hun}. In this (high) temperature regime, the magnitude of
the scaling parameter was suggested to be related to the
single-ion Kondo energy. The fact that the magnetoresistance could
be used to demarcate the crossover from an incoherent to coherent
Kondo scattering regime, and the observation that our scaling is
valid only within the coherent Kondo regime are in line with a two
fluid description of the Kondo lattice \cite{nak2}: It was shown
that while the scaling in the incoherent regime is influenced by
the single-ion Kondo energy scale, the scaling in the coherent
regime (as is observed in our data) should be dictated by the
intersite coupling energy. The latter is a measure of the
effective RKKY interaction between localized moments and is known
to play a crucial role, in addition to crystal electric field
splitting and the single-ion Kondo scale \cite{aep}. The above
mentioned fact of an effectively weaker antiferromagnetic
interaction in CeIrIn$_5$ in comparison to its Co counterpart is
consequently manifested in the value of the intersite coupling
energy scale $T^*$ ($\approx\! 20$ K and $\approx\! 45$ K in the
Ir and Co systems, respectively). The electronic ground state in
these systems is clearly dictated by the strength of $T^*$ in
relation to the single ion Kondo scale ($T_K$). This is evident
from the fact that CeRhIn$_5$ (which has a magnetically ordered
ground state) has a large value ($\approx\! 130$) of $T^* / T_K$
\cite{yan}. This ratio is found to be progressively smaller in the
Co and Ir counterparts ($\approx\! 25$ and $\approx\! 7$,
respectively). This qualitatively supports the results of our
scaling analysis, and is also in agreement with the
low-temperature phase diagrams reported for these systems.

In summary, we analyzed the normal-state magnetotransport in the
heavy-fermion superconductor CeIrIn$_5$. In the temperature regime
investigated here, the Hall and transverse conductivities do not
appear to be uniquely correlated. This is in marked contrast to
earlier observations on CeCoIn$_5$. The Hall coefficient $|R_H|$
is seen to satisfy a single parameter scaling of the form $R_H = f
[H / (a + b \, T^c)]$ in the coherent Kondo regime. This fitting
not only appears to be sensitive towards the fact that there is no
QPT in the vicinity of superconductivity in this system, but also
reflects (in an indirect fashion) the intrinsic nature of the
disparate bulk and resistive superconducting transitions.\\[0.2cm]

\hspace*{-0.3cm}{\bf IV. ACKNOWLEDGEMENTS}\\[0.3cm]
S.N. was supported by the Alexander von Humboldt Foundation. Work
at Dresden was supported by the EC (CoMePhS 517039) and the DFG
(Forschergruppe 960). Work at Los Alamos was performed under the
auspices of the U.S. Department of Energy/Office of Science.

\end{document}